\documentclass[journal]{IEEEtran}
\usepackage{cite}

\ifCLASSINFOpdf
   \usepackage[pdftex]{graphicx}
\else
   \usepackage[dvips]{graphicx}
\fi
\usepackage{booktabs}
\usepackage{multirow}
\usepackage{multicol}
\usepackage{algorithm}

\usepackage{amsmath}
\usepackage{algorithmic}

\usepackage{array}
\begin{document}
%
\title{Strongly Connected Topology Model and Confirmation-based Propagation Method for Cross-chain Interaction}
%
%
%

\author{Hong~Su
\IEEEcompsocitemizethanks{
\IEEEcompsocthanksitem This work has been submitted to the IEEE for possible publication. Copyright may be transferred without notice, after which this version may no longer be accessible.
\IEEEcompsocthanksitem H. Su is with College of Computer Science, Sichuan University, Chengdu, China.\protect\\
}
\thanks{}}

%
%

\markboth{Journal of \LaTeX\ Class Files,~Vol.~14, No.~8, August~2015}%
{Shell \MakeLowercase{\textit{et al.}}: Bare Demo of IEEEtran.cls for IEEE Journals}
%



\maketitle

\begin{abstract}
Cross-chain interaction is among different blockchains. When the number of blockchains increases, it is difficult for blockchains to form a single star topology or a fully connected topology. Meanwhile, different from legacy networks, the propagation method is required to keep the data validity. Thus, the blockchain topology and associated propagation methods are two key issues, which should be ensured during the propagation. In this paper, we first propose the confirmation-based propagation method to keep the validity of the cross-chain data. The confirmation method is to seal the cross-chain transaction synchronized from other blockchains. Second, we point out that a valid topology requires blockchains to be strongly connected. With this requirement, we propose several topologies, which match different connection scenarios. The simulation results show that the proposed methods can securely propagate the cross-chain data and the connection way is flexible.
\end{abstract}


%
\IEEEpeerreviewmaketitle

\section{Introduction}
\IEEEPARstart{B}{lockchain} has been successfully applied to Bitcoin initially \cite{pre_1}, and then applied to more digital currency areas \cite{pre_2}\cite{pre_3}. Besides the digital currency, blockchain has also been used in other aspects \cite{pre_4}, including exchanging digital assets \cite{pre_5}, and improving the supply or delivery system \cite{pre_7}\cite{pre_8}. 

To interact with different blockchains, cross-chain technologies \cite{pre_14}\cite{pre_16}\cite{pre_17} are introduced, which either transfers digital assets or shares transaction status among blockchains. Typical cross-chain interaction methods include the notary mechanism, sidechain, and hash-locking \cite{pre_18}.

Cross-chain interaction can occur among more than two blockchains \cite{pre_19}\cite{pre_23}. Take an e-shopping system as an example. This system includes three sub-systems, an electronic marketplace, a payment system, and a delivery system, which are based on different blockchains. A customer uses digital coins to buy goods from the electronic marketplace; the seller delivers the goods, and the shipping information of the goods is recorded in the delivery system. When the delivery system shows the goods have been received by the customer, the digital coin pre-paid by the customer is given to the seller. Figure 1 shows the cross-chain interaction of the three blockchain systems. This is interaction across three blockchains.

In an exchange, cross-chain transactions have dependencies on others. In the above example, the transaction (notated as $tx\_p$) in the payment blockchain depends on the transaction (notated as $tx\_em$) on the electronic marketplace blockchain and the transaction (notated as $tx\_d$) on the delivery blockchain. Transactions of $tx\_em$ and $tx\_d$ also depend on transaction $tx\_p$.

\begin{figure}[!t]
\centering
\includegraphics[width=2.2in]{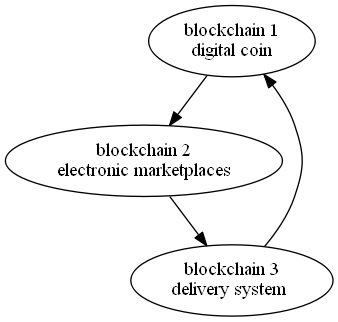}
\caption{An e-shopping system with three blockchains.}
\label{introduction_example}
\end{figure}

As cross-chain transactions are dependent, it requires to share the cross-chain data among blockchains. Blockchain 1 gives the payment to the seller when blockchain 1 gets the information that (1) there is an order in blockchain 2 between the customer and the seller, and that (2) the corresponding goods have been delivered to the customer. Then, associated transactions need to share among those blockchains. One way to share transaction information is to propagate them among different blockchains \cite{pre_19}\cite{pre_22}\cite{pre_23}.

When the propagation is among several blockchains, there are two aspects that should be considered. They aim to propagate the cross-chain transactions to other associated blockchains effectively. 

(1) The blockchain topology. It is the way how blockchains connect (instead of how nodes connect within a blockchain). The blockchain topology should be a flexible one. Flexibility refers to the amount of work that is required to add or change a blockchain in a topology. Less work is required, more flexible it is.

Currently, the blockchain topology is not obviously explored. There are two major kinds of blockchain connection ways. The first one is among two special blockchains (such as the pegged sidechain), which forms either a mutually connected topology or a star topology. The second one is to exchange data among blockchains by a blockchain router \cite{pre_22}\cite{pre_23}, which connects other blockchains.

However, there are issues for the star (router) topology. In work \cite{pre_22}, a single node from each blockchain forms a new router blockchain to communicate with other blockchains, which is flexible for scalability, while it has some potential issues. The node, which represents a blockchain, is the trust center for that blockchain. However, we cannot trust a single node, as the consensus algorithm does not depend on one node.

\begin{figure}[!t]
    \centering
    \includegraphics[width=2.5in]{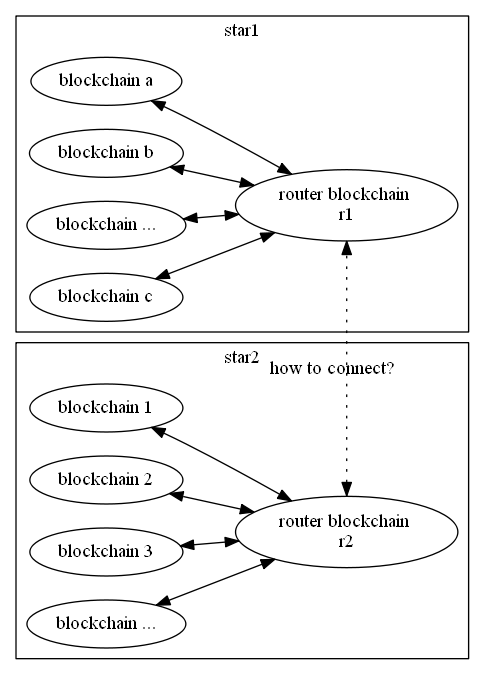}
    \caption{Star (centered) blockchain topology lacks scalability.}
    \label{2ctoconnect}
\end{figure}

Work \cite{pre_23} proposes to confirm cross-chain transactions during the propagation process, while it may have the issue of scalability. Suppose, there are two blockchain networks (star1 and star2), shown in Figure \ref{2ctoconnect}. In star1, all blockchains connect to router blockchain r1; and similarly, in star2, all blockchains connect to router blockchain r2. Which topology should be chosen when trying to connect star1 and star2? If the star topology is still chosen (such as r1 as the center), there are still two obvious issues. One is that all blockchains, which connect to r2, have to disconnect from r2 and switch to r1. It is a big change as all r2-connected blockchains are involved. The second is that all cross-chain data go through router blockchain r1, whose network flow increases. If the network burden of r1 is heavy enough, r1 has high network overload risk when r2-connected blockchains connect to it. A better way is to adopt a flexible topology to connect those two blockchain networks.

    (2) How to keep the data validity during the propagation. If one blockchain only propagates the data without keeping the data validity, its successive blockchains cannot trust this data, and it has no difference to propagate data on legacy networks.

    This paper proposes a cross-chain solution concerning both the scalability and data validation. We propose a strongly connect topology model, which guides to choose the possible topologies instead of providing a fixed topology. Based on the topology, we propose the method to keep the data validity between connected blockchains. The main contributions of this paper are as follows.

    (1) We point out that a blockchain topology requires corresponding blockchains to form a strongly connected graph. The strongly connected topology provides the path for the cross-chain data to propagate among associated blockchains. Based on this requirement, we propose some topologies to choose from when blockchains connect.

    (2) We propose a flexible propagation method to keep the validity of cross-chain data and to transform its data format. This method is done between directly connected neighbors and iterates to other neighbors. It only requires to transform and keep the validity of the cross-chain data from its neighbor.

    The rest of the paper is organized as follows.Section 2 describes the topology model and several topology instances. Section 3 discusses the propagation method based on the blockchain topology. Section 4 shows the verification results. Section 5 describes cross-chain related work. And section 6 gives a brief conclusion of this paper.

\section{Strongly Connected Topology Model}
The topology of blockchains describes how one blockchain connects to other blockchains. The topology model aims to provide a flexible connection way.

We first describe the connection types between two blockchains as it can be used to analyze the blockchain topology among several blockchains.

\subsection{Assumption}
In this paper, we have the following three assumptions. Notice, the notation of 'blockchain' can be the name of a blockchain or its chain of blocks; we use the term 'chain of blocks' as the chained blocks of a blockchain in confusing situations.

(1) Associated blockchains are based on P2P methods. Nodes of a blockchain connect to a certain number of other nodes to get and receive blockchain data (transactions and blocks) of this blockchain.

(2) Chain of blocks of a blockchain can be continuously got by other nodes, which may belong to the same blockchain or other blockchain(s). Public blockchains match this condition. For permissioned blockchain, it is required to add permissions for nodes of associated blockchain to access the blockchain data (no need to grant the access to do its consensus).

(3) If a blockchain (suppose blockchain $A$) wants to connect to another blockchain (suppose blockchain $B$), blockchain $A$ knows how to verify whether a block is correct or not by the consensus algorithm of blockchain $B$. This helps nodes of blockchain $A$ to make sure that information from $B$ is correct and keep the data validity. The simplest implementation way is that nodes of $A$ act as the verification nodes (instead of mining nodes) of $B$.

\subsection{Connection Types between Two Blockchains}
The connection of two blockchains is divided into three types: directly connected, indirectly connected, or not connected. We only describe the first two cases in which two blockchains can exchange transaction information.

\subsubsection{Directly Connected} \label{Directly_Connected}
(Blockchain) $A$ directly connects to (blockchain) $B$ when they match the following conditions. 

(1) Synchronization blockchain data from the directly connected blockchain. Nodes of $A$ get blockchain data (chain of blocks and transactions) from nodes of $B$ by P2P method. $A$ has a copy of the whole blockchain data of $B$. We also say blockchain $A$ gets information from $B$.

(2) Verification of blockchain data in the directly connected blockchain. Nodes of $A$ follows $B$'s consensus algorithm to choose the main chain of $B$ when there are forks in $B$, and to verify whether a specific transaction is correct or not.

(3) Sealing cross-chain transactions into its own blockchain. If the blockchain data of $B$ contains a cross-chain transaction, miners of $A$ try to transform and seal it. It is the same as to mine transactions of blockchain $A$, except that transactions from $B$ are marked as cross-chain transactions. An extra reward for the miner is used to encourage miners to synchronize and seal the cross-chain transactions from associated blockchains.

\begin{figure}[htp]
    \centering
    \includegraphics[width=3in]{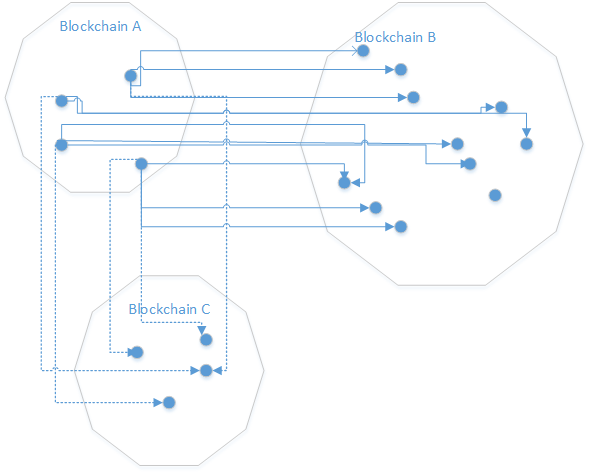}
    \caption{Directly connected blockchains. Blockchain $A$ directly connects to blockchain $B$ and $C$, in which nodes of blockchain $A$ connect to certain number of nodes of blockchain $B$ (solid arrow) and blockchain $C$ (dotted arrow).}
    \label{direct_connection}
\end{figure}

\begin{figure}[htp]
    \centering
    \includegraphics[width=2.3in]{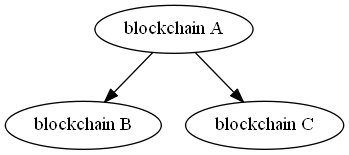}
    \caption{Simplified diagram for directly connected blockchains.}
    \label{direct_connection_simple}
\end{figure}

One direct connection has its direction. $A$ directly connects to $B$ does not mean that $B$ also directly connects to $A$ if nodes of $B$ do not get the chain of blocks from nodes of $A$ or do not follow $A$'s consensus algorithm. Figure \ref{direct_connection} shows an example, in which $A$ directly connects to $B$ and $C$ while $B$ or $C$ does not directly connect to $A$. For simplification, Figure \ref{direct_connection} can be reformed into Figure \ref{direct_connection_simple}, in which we hide details of nodes.

\subsubsection{Indirectly Connected}
(Blockchain) $A$ indirectly connects to (blockchain) $B$ when it matches the following conditions.

(1) Nodes of $A$ does not get the blockchain data from nodes of blockchain $B$.

(2) Or nodes of $A$ does not follow the consensus of blockchain $B$ to validate the blockchain data of $B$.

(3) However, blockchain $A$ still connects to blockchain $B$. Nodes of blockchain $A$ gets cross-chain transactions of blockchain $B$ through other blockchains.

\subsection{Strongly Connected Topology}
To describe the topology of blockchains, we use the directed graph. If (blockchain) $A$ directly connects to (blockchain) $B$, we use a directed edge to stand for the relationship, pointing from $A$ to $B$. All edges consist of a directed graph.

\begin{figure}[htp]
    \centering
    \includegraphics[width=2.5in]{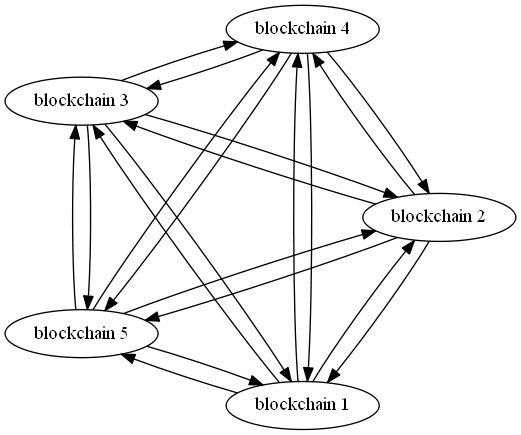}
    \caption{Fully connected topology example.}
    \label{mutli_dependence_all_connected}
\end{figure}

The essential characteristics of blockchain topology require connected blockchains to form a strongly connected graph, in which blockchain data can be exchanged. Figure \ref{mutli_dependence_all_connected} shows a blockchain topology with all blockchains directly connected.

Why the topology graph is strongly connected? We briefly prove this by reduction to absurdity. If the graph is not strongly connected and then at least one node (one blockchain) cannot be reached. Assume that node is node N. It means no other nodes can get blockchain data from N, and then cross-chain transactions on this blockchain cannot propagate to other blockchains. Thus, no blockchain should be unreachable. The topology graph of corresponding blockchains is a strongly connected graph. 

Now we describe some typical topology models.

Fully connected topology is a topology in which all blockchains are directly connected. Figure \ref{mutli_dependence_all_connected} is one example; each blockchain connects to all other blockchains. However, the connection is complicated, and the number of connections increases when the number of blockchain increases. Suppose, each node of a blockchain needs to connect at least $k$ nodes of another blockchain. If there are $j$ blockchains, each node needs to connect $k*(j-1)$ nodes of other blockchains.

\begin{figure}[htp]
    \centering
    \includegraphics[width=3.3in]{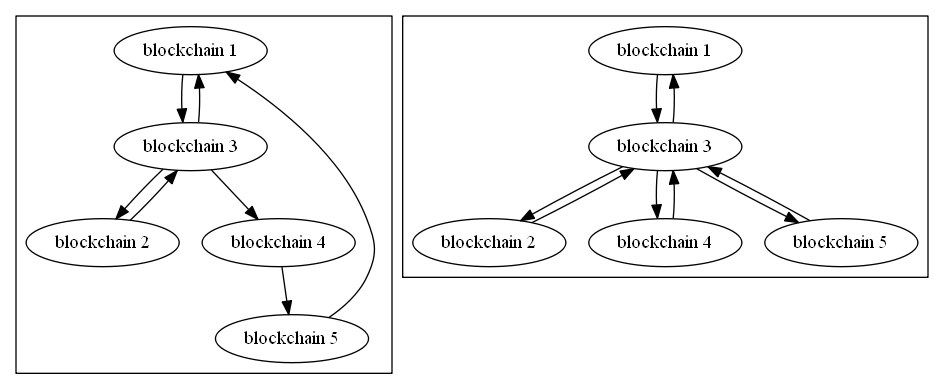}
    \caption{Star topology example}
    \label{mutli_dependence_star}
\end{figure}

Star topology is a topology in which associated blockchains communicate through a center blockchain (the router blockchain). This kind of topology is also called the router topology. Figure \ref{mutli_dependence_star} shows an example. In the left part, blockchain 3 acts as the center node; blockchain 1 and blockchain 2 communicate through blockchain 3; in the right part, blockchain 3 acts as the center nodes of all other blockchains.

The star topology has a center blockchain, which may cause some potential issues. (1) The center blockchain has a heavy communication burden; (2) it is relatively big work to merge two star topologies as discussed in the introduction section.

Now we describe another topology of blockchains which only requires one blockchain directly connects to one another blockchain, called the ring topology.

\subsection{Ring Topology – An kind of Strongly Connected Topology}
As it only requires the blockchain topology to form a strongly connected graph, the fully connected topology can be simplified. One simplified method is that a blockchain only directly connects to another one. This kind of topology is called the ring topology.

\begin{figure}[htp]
    \centering
    \includegraphics[width=2.5in]{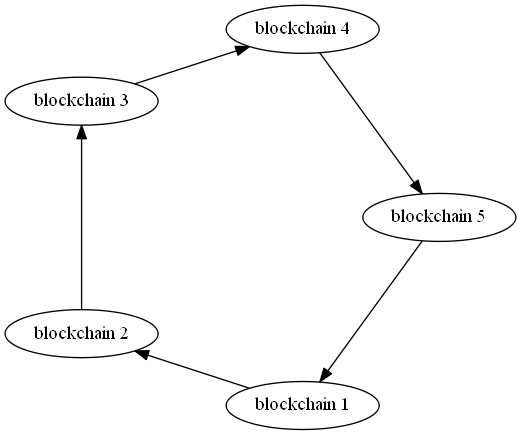}
    \caption{Ring topology example.}
    \label{mutli_dependence}
\end{figure}

In the graph of the ring topology, each blockchain (as a node) has an outgoing edge (pointing to other nodes) and in-coming edge (pointing to this node). Both edges are necessary. If a node has only one of the two edges, it cannot exchange blockchain data with other blockchains. In this way, all nodes form a strongly connected graph. Figure \ref{mutli_dependence} shows an example, in which each blockchain connects to another one and those five blockchains form a ring topology.

\subsubsection{Bridge Blockchain}
In a blockchain topology, some blockchains only help to propagate the cross-chain data for other indirectly connected blockchains. Those blockchains are called bridge blockchains. They do not have their own cross-chain transactions and only participate in the propagation of the cross-chain interaction. 

\begin{figure}[htp]
    \centering
    \includegraphics[width=2.6in]{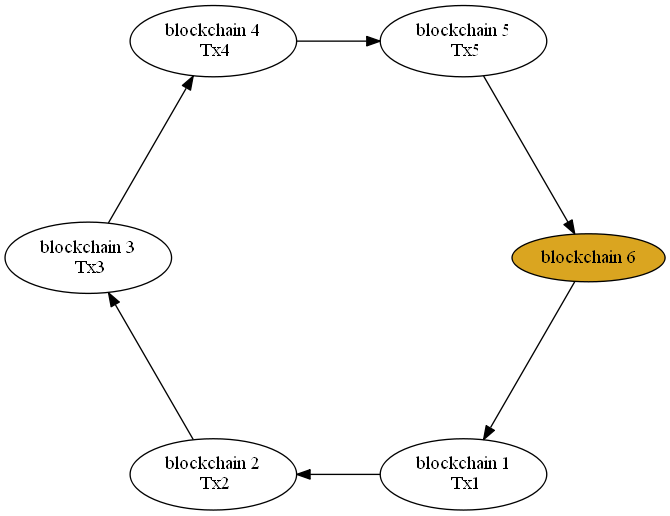}
    \caption{An example of the bridge blockchain (blockchain 6).}
    \label{mutli_dependence_with_Tx_bridge}
\end{figure}

Figure \ref{mutli_dependence_with_Tx_bridge} shows one example, in which blockchain from 1 to 5 have their associated transactions (named from $Tx1$ to $Tx5$ separately). There is no cross-chain transactions from blockchain 6. It acts as the bridge blockchain, which makes other blockchains strongly connected.

\subsection{Connecting to New Blockchain}
To form or join a blockchain topology, a blockchain is required to connect to other blockchain(s). We propose a special kind of transaction (the topology proposal transaction) that is used to initialize the proposal for the connection to another blockchain. It is sent from specific accounts, namely topology selection accounts. Those accounts can be pre-configured in the source code or match some conditions (such as accounts who have the top maximum number of assets). 

After a topology proposal transaction is put into the blockchain, topology selection accounts send another kind of transaction (the agreement transaction) if they agree with the proposal. Agreement transactions aim to avoid error sending of the topology proposal transaction. Certain portion instead of all those accounts are required to agree, as some topology selection accounts may be inactive. 

There are some limitations to the connection proposal. The maximum number of directly connected blockchains should be limited, aiming to restrict the number of external nodes to connect. Meanwhile, some aspects of a target blockchain are evaluated for the connection; if the target blockchain does not match, the new connection is still not carried out.

\section{Confirmation-based Propagation Method}
In this section, we describe the proposed cross-chain data propagation method, the confirmation-based propagation method (CBT), which aims to keep data validity during the propagation.

\subsection{Confirmation-based Propagation Process}
Different from legacy networks, data validity should be confirmed in its directly connected blockchain. If one blockchain only propagates the data without confirmation, its successive blockchains cannot trust this data, and it does not have much difference to propagate the transaction data on a legacy network.

The confirmation process is similar to the process for an internal transaction. The validity of an internal transaction is kept by sealing it to the blockchain with the consensus algorithm. Then for a cross-chain transaction, it can still be achieved by sealing it into associated blockchains. We select the ring topology as an example to analyze.

After a blockchain gets cross-chain transactions from its directly connected blockchain, its miners process to confirm those transactions into its own blockchain. This is called the confirmation-based propagation process. It can be divided into two steps. In the first step, corresponding nodes transform the format of cross-chain transactions, which will be described in “Format Transformation of Cross-chain Data”. In the second step, miners of target blockchain try to seal transformed cross-chain transactions to its own blockchain.

Algorithm \ref{alg_re_se_fr} is the algorithm for the above process (the synchronization process is also included), which is between two directly connected blockchains. Line \ref{alg_sync} synchronizes the blockchain data, and line \ref{alg_check} checks the validation of the blockchain data. Lines from \ref{alg_get_new_list} to \ref{alg_transform} are the second steps, in which new cross-chain transactions are selected and transformed. Lines from \ref{alg_judge_mine} to \ref{alg_all_done} are the third steps, in which the cross-chain data is sent to miner for sealing.

\begin{algorithm}[htb]
  \caption{ To retrieve and seal the cross-chain data from directly connected blockchain. }
  \label{alg_re_se_fr}
  \begin{algorithmic}[1] 
      \REQUIRE ~~\\
      blockchain $dcb$ is the directly connected blockchain of the current blockchain. Transactions are sealed in blocks for simplification.
      \ENSURE ~~\\
      cross-chain transactions in $dcb$ is sealed into the current blockchain

      \STATE 		  $copiedBlockchain$ = synchronize chain of blocks from directly connected blockchain $dcb$ \label{alg_sync}
      \IF         {$copiedBlockchain$ is invalid} \label{alg_check}
        \STATE    return;
      \ENDIF
      \STATE 		  $transactionList$ = get transaction list from new blocks of $copiedBlockchain$ \label{alg_get_new_list}
      \STATE 		  for each transaction $t$ in $transactionList$ 
      \IF 		    {$t$ instanceof Cross-chainTransaction}
        \STATE 				//transform different data format between directly connected blockchains 
        \STATE 				$newTransaction$ = transform the format of $t$; \label{alg_transform}
        \STATE 				// other nodes may seal first, then we judge whether it has already been put into this blockchain or not \label{alg_judge_mine}
        \IF 				  {$newTransaction$ does not exist in current blockchain}
          \STATE 					// Let miner to seal this transaction
          \IF         {mining node}
            \STATE    try to seal	$newTransaction$ into blockchain;
          \ENDIF
        \ENDIF
      \ENDIF \label{alg_all_done}
  \end{algorithmic}
\end{algorithm}

\begin{figure}[htp]
    \centering
    \includegraphics[width=3in]{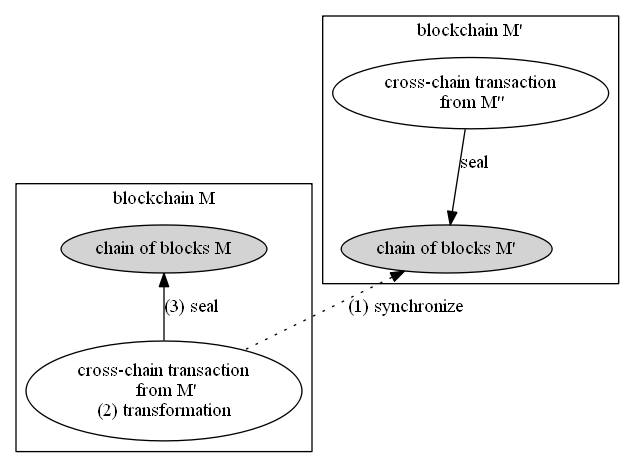}
    \caption{The process to synchronize and seal cross-chain transaction from directly connected blockchain.}
    \label{duplicated_ledger_v1}
\end{figure}

  We use Figure \ref{duplicated_ledger_v1} to explain this process, in which blockchain M directly connects to blockchain M+1. Nodes of blockchain M synchronize the blockchain data from their connected nodes of blockchain M+1 periodically – arrow marked with '(1) synchronize'. The cross-chain transaction is selected, and its format is transformed into the format of blockchain M – marked with '(2) transformation'. Miners of blockchain M try to seal this transaction into its own blockchain – marked with '(3) seal'.

\subsubsection{Propagation Between Indirectly Connected Blockchains}
The above is the propagation process between directly connected blockchains, which iterates to propagate data to indirectly connected blockchains.

We use a function ($Prop$) to describe the iterative propagation process. This function is the abstract of propagation steps between directly connected blockchains $X$ and $Y$. It indicates that transaction $CrTx\_X$ on blockchain $X$ is propagated to blockchain $Y$ as transaction $CrTx\_Y$, referring to \eqref{prop}.

\begin{equation} \label{prop}
  \begin{array}{l}
    CrTx\_Y = Prop(CrTx\_X)\;\;\\
    {\rm{, where\ }}CrTx\_X{\rm{\ is\ a\ cross-chain\ data\ in\ blockchain\ }}X\\
    {\rm{,and\ }}CrTx\_Y{\rm{\ is\ the\ cross-chain\ data\ in\ blockchain\ }}Y\\
    {\rm{after\ one\ propagation.}}\;\;\;
    \end{array}
\end{equation}

Now suppose there are three blockchains, $M$-1, $M$, and $M$+1. $M$-1 directly connects to $M$, and $M$ directly connects to $M$+1. Then we get propagation description as in \eqref{mtomp1} \eqref{mtoms1} \eqref{allProp}.

\begin{equation} \label{mtomp1}
  CrTx\_M = Prop\left( {CrTx\_M+1} \right)
\end{equation}

\begin{equation} \label{mtoms1}
  CrTx\_M - 1 = Prop(CrTx\_M)
\end{equation}

\begin{align}  \label{allProp}
  CrTx\_M - 1  &= Prop\left( {CrTx\_M} \right){\rm{ }} \\ 
               &= Prop(Prop(CrTx\_M + 1)) \notag
\end{align}

From \eqref{allProp}, we see the cross-chain data $CrTx\_M$+1 in $M$+1 has been propagated to its indirectly connected neighbor $M$-1. Data validity is kept during this propagation process, and it has even been enhanced as all associated blockchains have sealed the cross-chain data.

For any indirectly connected blockchain $M$ and $N$, there is at least one directed path from $M$ to $N$. $CrTx\_M$ propagates along one of those paths to $N$. Suppose the path is $M->…->N$$+1->N$, then we get the propagation as in \eqref{proppath}.

\begin{align}
  CrTx\_N &= Pro\left( {CrTx\_M} \right){\rm{ }} \notag \\
          &= Pro\left( Pro\left( {CrTx\_M} \right) \right){\rm{ }} \label{proppath} \\
          &= {\rm{ }}Pro(...(Pro(CrTx\_M))) \notag
\end{align}

\begin{figure}[htb]
  \includegraphics[width=3in]{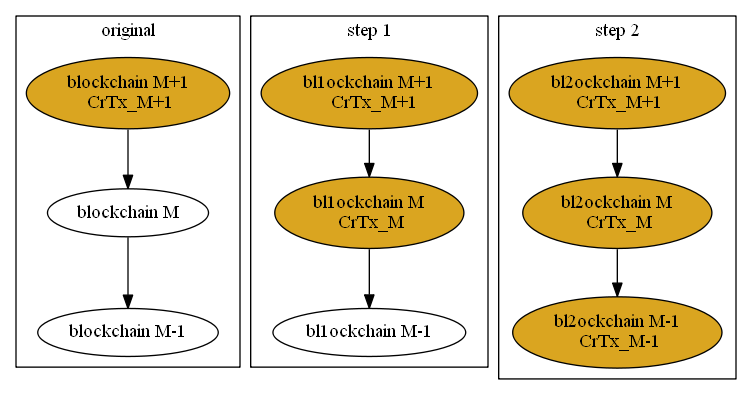}
  \caption{The transaction propagation to indirectly connected blockchains.}
  \label{message_propagation}
\end{figure}

This process is shown in Figure \ref{message_propagation}. Cross-chain transaction ($CrTx\_M$+1) is put to blockchain $M$+1 (left part of Figure \ref{message_propagation}). Then this transaction data propagates to its directly connected neighbor $M$ (the middle part of Figure \ref{message_propagation}). At last, blockchain $M$ propagates this transaction to its directly connected neighbor $M$-1 (right part of Figure \ref{message_propagation}). Thus, $CrTx\_M$+1 propagates from blockchain $M$+1 to its indirectly connected blockchain $M$-1.

\subsection{Format Transformation of Cross-chain Data}
The data structures of transactions may be different among blockchains. If we transform data format between all possible blockchain pairs, the number of transformations is the square of the number of blockchains.

There are two possible ways to handle the transformation of the cross-chain data. One is to unify the format of the cross transaction within all blockchains – the unified way; another one is to translate the cross-chain date of directly connected blockchains – the translation way. 

There are some related works on the unified way \cite{pre_23}. In the unified way, each blockchain has a unified transaction data structure instead of its own transaction structure. While it still needs to transform the data format between the cross-chain transaction and the local transaction; it requires all blockchains to adopt the unified cross-chain transaction data format. In this paper, we introduce the translation way between directly connected blockchains.
\subsubsection{Translation Way}
\begin{figure}[htb]
  \includegraphics[width=3.5in]{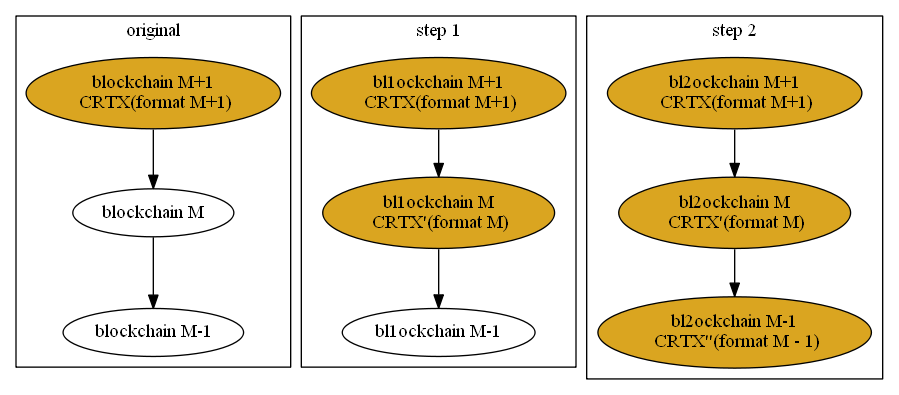}
  \caption{Blockchain data transformation process.}
  \label{message_propagation_protocol}
\end{figure}

In this way, we should consider how to eliminate the number of the required transformations. In our proposed way, each blockchain transforms the cross-chain data of directly connected blockchains. This process is repeated in all directly connected blockchains. It reduces transformations required for indirectly connected blockchains. Figure \ref{message_propagation_protocol} shows this process.

This process is similar to the propagation process and is done in the propagation steps. In Figure \ref{message_propagation_protocol}, one transaction ($CRTX$) of blockchain $M$+1 is in original format 'M+1'; its directly connected neighbor (blockchain $M$) gets this transaction information and translates it from format 'M+1' to format 'M' ($CRTX'$) which is the format used in blockchain $M$. Blockchain $M$-1 knows how to transform format 'M-1' and it translates format from 'M' to 'M-1'. 

From the transformation point of view, we say blockchain $A$ directly connects to blockchain $B$ refers to that (1) blockchain $A$ translates the transaction format of blockchain $B$, and (2) seals the transformed transaction into its own blockchain.
We introduce a transformation function $Transf$ to give a formal description of the transformation process. The transformation from $f_n$ to $f_m$ is expressed as \eqref{eq_transform}, which is between directly connected blockchains.

\begin{equation} \label{eq_transform}
  \begin{array}{l}
    f_m = Transf_{n\_m}(f_n)\\
    {\rm{, where\;}}f_n{\rm{\ is\ the\ original\ format\ and\ }}f_m{\rm{\;is\ the\ format}} \\
    {\rm{after\ transformation;\;}}Transf_{n\_m}{\rm{\ is\ the\ function\ to}} \\
    {\rm{transform\ transaction\ from\ format\; }}f_n{\rm{\ to\ }}f_m{\rm{.}}
    \end{array}
\end{equation}

Now we begin to describe the transformation process in the propagation path. Suppose there is a blockchain sequence in a ring topology as in \eqref{bsdrt}.

\begin{equation} \label{bsdrt}
  [bc_1,bc_2,...,bc_n]
\end{equation}

Then the associated formats are also in a sequence, which is in formula \eqref{bsdrtft}.

\begin{equation} \label{bsdrtft}
  \begin{array}{c}
  [f_1,f_2,...,f_n] \\
  {\rm{,where\ }}f_n{\rm{\ is\ the\ blockchain\ data\ format\ in\ blockchain\ }}n{\rm{.}}
  \end{array}
\end{equation}

The transformation process is described as in \eqref{cafrttrans}.

\begin{equation} \label{cafrttrans}
  \begin{array}{l}
    f_1 = Transf_{n\_1}(f_n)\\
    f_2 = Transf_{1\_2}(f_1)\\
    ...\\
    f_n = Transf_{n - 1\_n}(f_{n - 1})
    \end{array}
\end{equation}

In this transformation sequence, we see each blockchain only needs to know how to transform the data of the directly connected blockchain and has only one transformation function $Transf_{n\_m}$. The total number of the transformation is $n$ which is linear to the number of blockchains while the data format can be transformed between any blockchain pairs.

\subsection{Security Analysis}
In the propagation process, there are two security enhancement processes for the directly connected blockchains. Suppose (blockchain) $A$ directly connects to (blockchain) $B$. First, nodes of $A$ check the validity of cross-chain transactions from $B$ by the consensus algorithm of $B$. And then miners of $A$ seal those cross-chain transactions by its own consensus algorithm. Then a cross-chain transaction of $B$ is confirmed twice. The final consensus process ($consensus\_f$) can be seen as in \eqref{twice_enhancement}.

\begin{equation} \label{twice_enhancement}
  \begin{array}{l}
  consensus\_f = consensus\_A(consensus\_B(Tx)) \\
  {\rm{,where\ }}Tx{\rm{\ is\ a\ cross-chain\ transaction,\ }}consensus\_B \\
  {\rm{and\ }} consensus\_A{\rm{\ are\ consensus\ algorithm\ of\ blockchain\ }}\\
  B{\rm{\ and\ }}A{\rm{\ respectively.}}
  \end{array}
\end{equation}

Then possible combinations of PoS and PoW are, $PoS(PoW(Tx))$, $PoW(PoW(Tx))$, $PoW(PoS(Tx))$, and $PoS(PoS(Tx))$. 

Suppose, $p_i$ is the possibility to fake a transaction on blockchain $i$. Thus, a successful fake of a cross-chain transaction requires to fake associated transactions in all associated blockchains. Then the possibility ($pb$) to fake a cross-chain transaction is as in \eqref{fakeAllBC}.

\begin{equation} \label{fakeAllBC}
pb = p_1*…*p_i*…*p_n
\end{equation}

From \eqref{fakeAllBC}, we know the possibility to fake a cross-chain transaction in associated blockchains is far less than the possibility to fake in a blockchain. This can also be understood by the propagation process of the cross-chain transaction. Suppose a node $n_m$ on blockchain $m$ fetches the data of a transaction $tx_n$ on blockchain $n$, and seals it as a copy $tx_m$ on blockchain $m$. Their validation methods are independent. $tx_m$ is validated by the consensus algorithm of blockchain $m$, and $tx_n$ is validated by the consensus algorithm of blockchain $n$. They are independent and then an adversary has to break two systems to achieve the cheating purpose. When more copies of this transaction have been sealed in other indirectly connected blockchains, it is even more difficult to break all corresponding blockchains.

If an adversary only breaks some blockchains, we can detect this as transaction copies are different among blockchains. We define a new variety ($pf$) as the possibility to detect that a cross-chain transaction has been changed. $pf$ is the opposite of the possibility that states of a transaction in all blockchains are the same, which includes cases that all blockchains have been changed or none of the blockchains has not been changed, as in \eqref{fakepf}.

\begin{equation} \label{fakepf}
  \begin{array}{l}
  pf = 1 - (1-p_1)(…)(1-p_i)(…)(1-p_n) – pb \\
  {\rm{,\ where\ }}1-p_i{\rm{\ is\ the\ possibility\ that\ a\ blockchain}} \\{\rm{has\ not\ been\ broken.}}
  \end{array}
\end{equation}

Then with the confirmation in its directly and indirectly connected blockchains, the inconsistency of data can be detected. This brings new features of the security of a cross-chain transaction. 

\section{Verification}
In this section, we describe and analyze the verification result, with the aim to show the flexibility of the proposed methods and how the data validity is kept during the propagation. It has three subsections. The first subsection describes the environment. The second subsection compares the strongly connected topology with the router topology (a star topology), and the third subsection verifies different connection way of the proposed methods.
\subsection{Verification Environment}
We set up four blockchains for cross-chain verifications, blockchain 1, 2, 3 and 4. Each blockchain has its own blockchain configuration, including its mining configuration, genesis block, and a unique blockchain number (blockchain identification number or simply blockchain id). Those blockchains are deployed on virtual machines. Different blockchains have different numbers of nodes, and resources of nodes are also different (CPU, 1, 2, or 4 vCPU; memory, 1G, 2G, 4G or 8G; hard disk, 32G or 64G).

The blockchain data has been divided into two types, internal blockchain data, and (external) cross-chain blockchain data. A blockchain has its own flag to identify a cross-chain transaction. This flag is used for miners of its directly connected blockchain to process cross-chain transactions.

The blockchain data transfers by the P2P method. We implement it by corresponding Java socket APIs. The nodes of directly connected blockchains are pre-configured in a file for simplification. Nodes of a blockchain (suppose $A$) connect to nodes of its directly connected blockchain (suppose $B$), and get the chain of blocks from $B$ periodically (every 1 second in our verifications). After a synchronization, miners of $A$ try to pick the cross-chain transactions, transform, and seal them into $A$’s chain of blocks. Then those transactions are in $A$’s own blockchain and are synchronized among internal peers of blockchain $A$.

Nodes of a blockchain act as verification nodes of its directly connected blockchain. Those nodes only run the verification code of the target blockchain, and they only do the consensus algorithm for its own blockchain.

There are two different consensus algorithms in the verifications, PoW and PoS. For the PoW algorithm, miners try to find a random number (nounce) to match the target difficulty, which requires that the hash of a block starts with a certain number of zeroes (6 zeros in the verifications). For simplification, the difficulty is fixed during the test process, and does not change to adjust the mining time.

For the PoS algorithm, we use (PoS) weight to select the account to seal a block. If an account has maximum weight, it seals the next block. PoS weight is a number and is calculated concerning the amount of an account’s asset. Its initial value is mapped linearly to the account of an account’s asset. At the construction stage, we initiate a certain number of accounts (16 accounts) with different numbers of assets. If one account has sealed a block, a certain amount of weight is substrated from it. To avoid possible mining conflicts, an account is only mounted to one node. The block is sealed on that node if the corresponding account is selected.

\subsection{Network Load Comparison among Router Topology and Ring Topology}
In this subsection, we compare the router topology (the star topology) and the proposed strongly connected topology (a ring topology). We adopt the network load to reflect the topology structure. The network flow refers to the amount of blockchain data exchanged among blockchains. In the router topology, all other blockchains connect to a router blockchain, and then their blockchain data go through the router blockchain. The ring topology is one kind of the strongly connected topology, in which the network load of a blockchain only goes to its directly connected blockchain.

\begin{figure}[htp]
  \centering
  \includegraphics[width=2in]{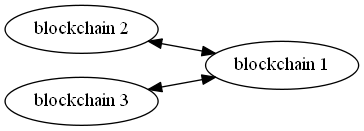}
  \caption{The router topology.}
  \label{1to2}
\end{figure}

\begin{figure}[htp]
  \centering
  \includegraphics[width=1.8in]{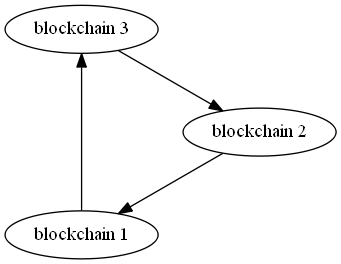}
  \caption{The ring topology.}
  \label{1to1to1}
\end{figure}

There are three blockchains in the verifications, blockchain 1, 2, and 3. In the router topology, blockchain 1 behaves as the router blockchain, which directly connects to blockchain 2 and 3. Blockchain 2  and 3 also directly connect to blockchain 1. The router topology is depicted in Figure \ref{1to2}. In the ring topology, blockchain 1, 2, and 3 directly connect to another one, shown in Figure \ref{1to1to1}. 

There are two subtype verifications, S1 and S2, with respect to different consensus algorithms of those blockchains. In S1, consensus algorithms of all blockchains are PoW. In S2, consensus algorithms of blockchain 1 and 2 are PoW, and that of blockchain 3 is PoS. We test different consensus to show cross-chain data can be also confirmed when the consensus algorithms are different. PoW and PoS are selected, as they are the most widely used consensus currently.

To send transactions continuously, we use scripts to send out cross-chain transactions, and one node of each blockchain is selected to run those scripts. The transaction sending rate is from 150 transactions per minute to 5000 transactions per minute. There are two ways to adjust the rate. (1) Manual way. This way aims to reduce the big difference of the network load among blockchains – to make sure there is no blockchain whose network flow is ten times of other blockchains. (2) Automatical way. The sending node automatically processes another transaction after the previous transaction is pre-processed; pre-processing includes to parse transaction data, to check the balance of the sender, and to send to other nodes. The higher the average pre-processing speed is, the more transactions one node process.

The network flow of a blockchain is the sum of the network flow from all nodes of a blockchain. If we only fetch network flow from partial of its nodes, it cannot reflect the outgoing network flow of the whole blockchain, as its directly connected blockchain may get the blockchain information from other nodes. A counter to measure the network flow of a node is added when socket APIs are invoked to send or receive packages to external blockchain nodes. The counter outputs network flow per second.

\begin{figure*}
  \includegraphics[width=7in]{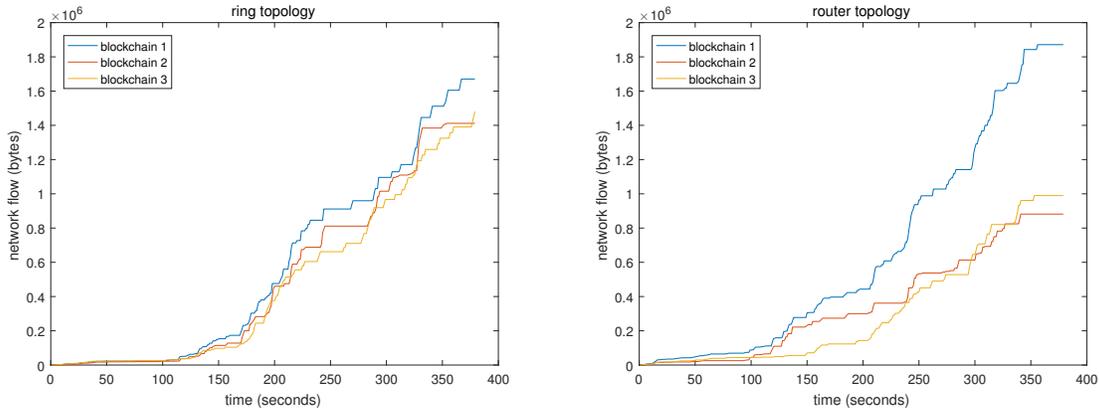}
  \caption{The network flow of S1. The left is for the router topology, and the right is for the ring topology.}
  \label{network_flow_s1}
\end{figure*}

The network flow of S1 is shown in Figure \ref{network_flow_s1}, in which the left part is for the router topology and the right is for the ring topology.

In the left part of Figure \ref{network_flow_s1} (the network flow of the ring topology), we see that those three blockchains have almost the same network flow over time. The exchange of cross-chain data occurs among directly connected blockchains instead of a centered blockchain. The network flow of each blockchain has no big difference as (1) the transaction sending rate is approximately the same, and (2) the mining difficulty is the same (6 zeros) for those three blockchains.

The right part of Figure \ref{network_flow_s1} shows the network flow of the router topology. Blockchain 1 has far more network flow than the other two blockchains do. For example, at 350 seconds, the network flow of blockchain 1 is 1,842,757 bytes per second, 881,557 and 961,200 bytes per second for blockchain 2 and blockchain 3 accordingly. This indicates that the network flow of the router blockchain is heavier than any of its connected blockchain. 

\begin{figure}[htp]
  \includegraphics[width=3in]{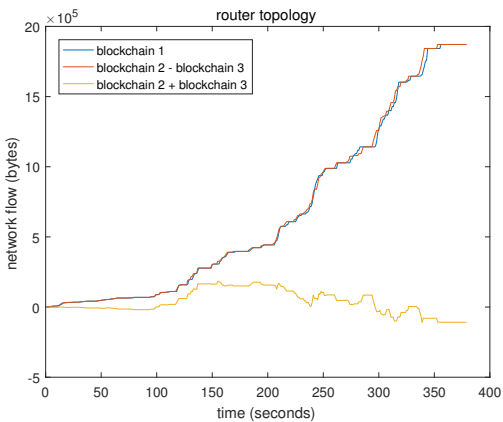}
  \caption{The network flow comparison of S1 for router topology. Notation of 'blockchain 2 - blockchain 3' means the difference of the network flow between blockchain 2 and blockchain 3; notation of 'blockchain 2 + blockchain 3' means their sum. We skip that for ring topology as each blockchain's network flow is almost the same.}
  \label{network_flow_s1_router_sum_and_difference}
\end{figure}

To further see the relationship of the router blockchain and its connected blockchains, we show the sum and the difference of network flow of router-blockchain-connected blockchains in Figure \ref{network_flow_s1_router_sum_and_difference}.

There are two interesting points in Figure \ref{network_flow_s1_router_sum_and_difference}. The first one is that the sum of the network flow of blockchain 2 and blockchain 3 is very close to that of the router blockchain (blockchain 1). At 360 seconds, the network flow of blockchain 1, blockchain 2, and blockchain 3 are 1,871,229bytes, 881,557 bytes , and 989,672 bytes in a second respectively. The network flow of blockchain 1 is equal to the sum of blockchain 2 and blockchain 3. It indicates that all cross-chain data are through router blockchain. We can predict that if the number of blockchains connecting to the router increase, the network flow increase greatly.

The second one is that the difference in the network flow among blockchain 2 and blockchain 3 is not obvious (referring to the curve of 'blockchain2-blockchain3' in Figure \ref{network_flow_s1_router_sum_and_difference}). The difference has no obvious tendency; there are minus or positive values of the difference interactively. This is due to that those two blockchains have the same mining difficulty (6 zeros) and the cross-chain transaction sending speed is the approximately same (around 3000 transactions per minute).

\begin{figure*}
  \includegraphics[width=7in]{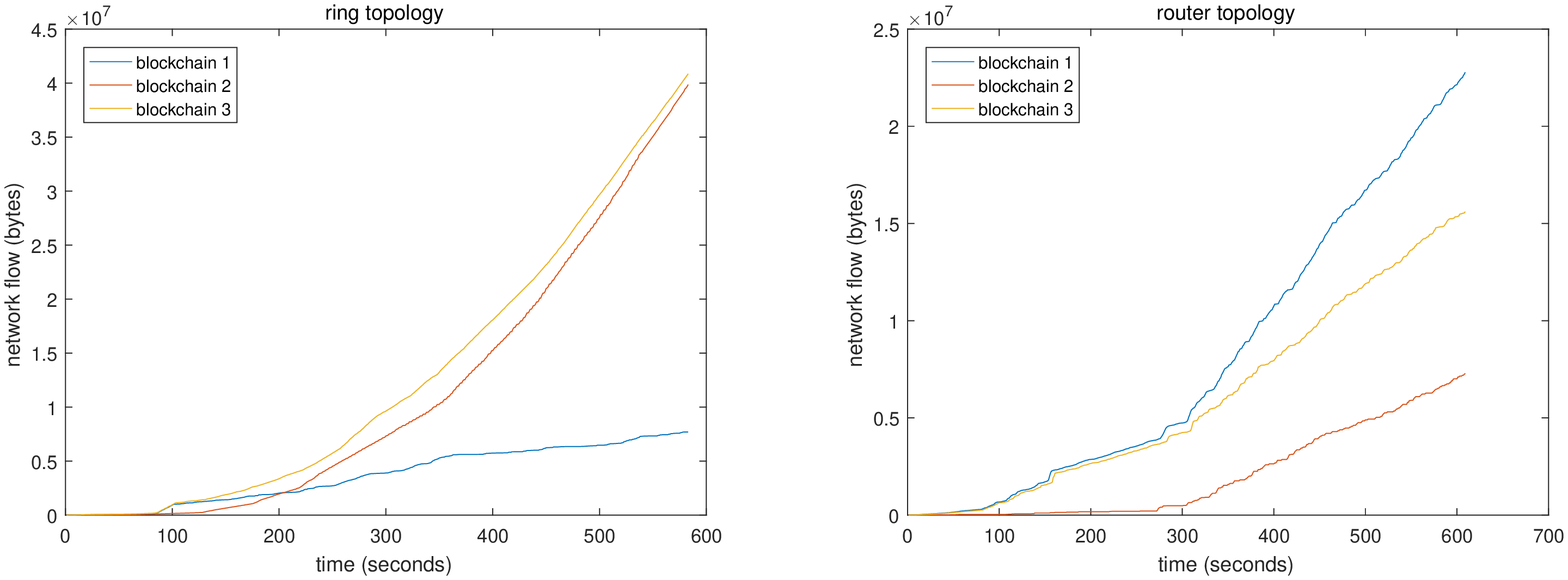}
  \caption{The network flow of S2. The left is for the ring topology, and the right is for the router topology.}
  \label{network_flow_s2}
\end{figure*}

The network flow of S2 in the ring topology and the router topology are shown in Figure \ref{network_flow_s2}. 

\begin{figure}[htp]
  \includegraphics[width=3in]{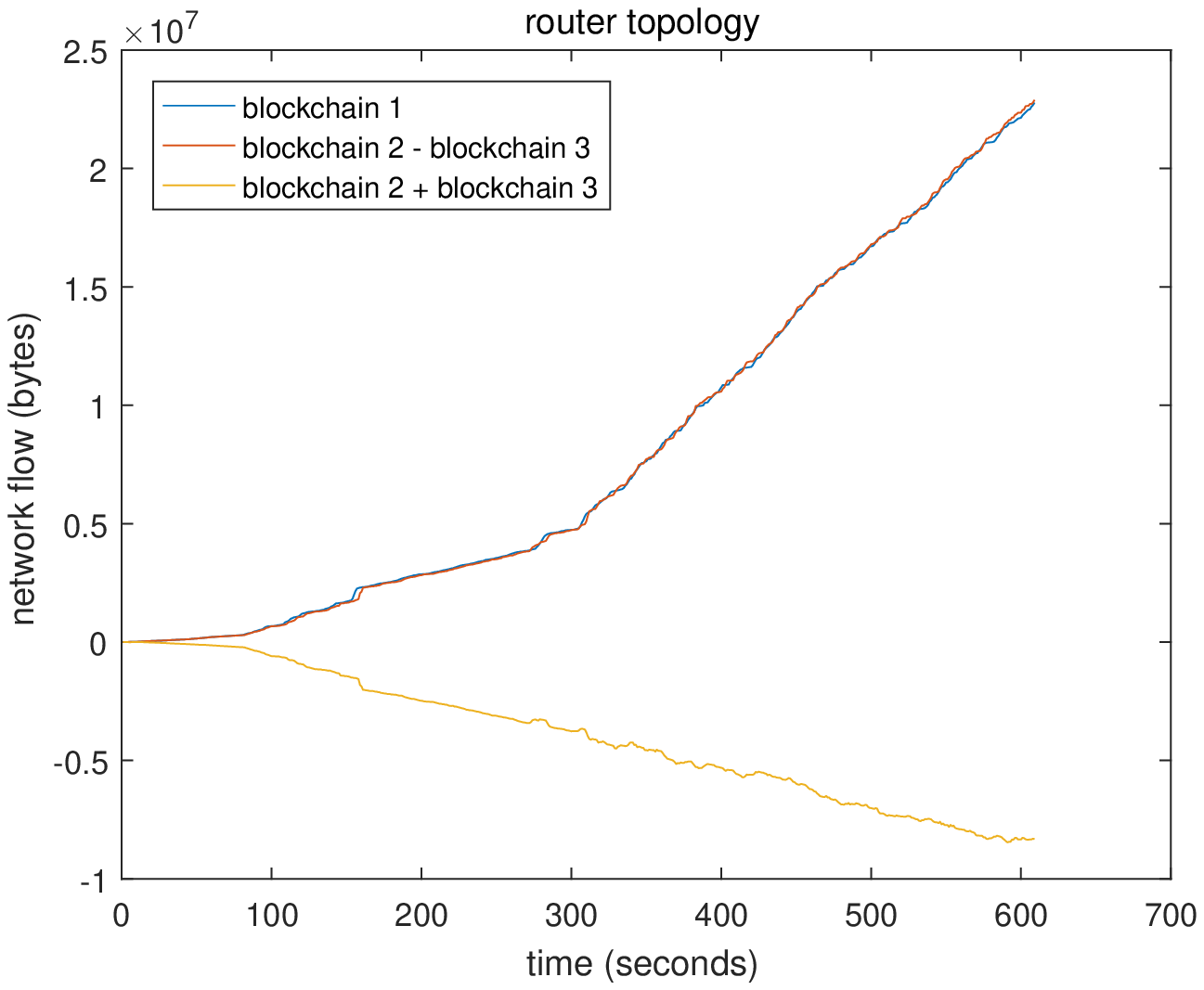}
  \caption{The network flow comparison of S2 for the router topology. Notation of 'blockchain 2 - blockchain 3' means the difference of the network flow between blockchain 2 and blockchain 3; notation of 'blockchain 2 + blockchain 3' means their sum. We skip that for ring topology as each blockchain's network flow is almost the same.}
  \label{network_flow_s2_router_sum_and_difference}
\end{figure}

\begin{figure}[htp]
  \includegraphics[width=2.4in]{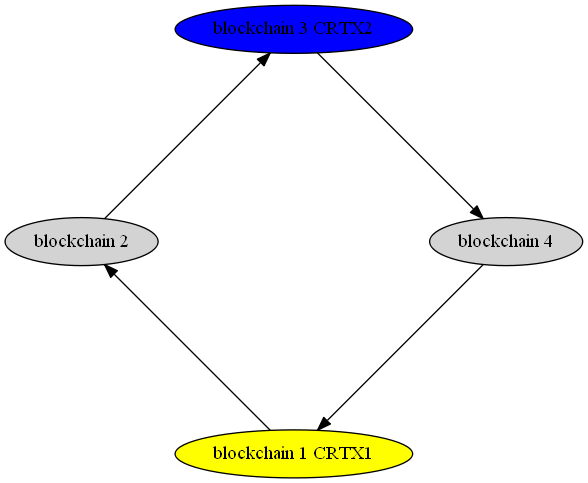}
  \caption{The topology of two indirectly connected blockchains.}
  \label{network1}
\end{figure}

The left part of Figure \ref{network_flow_s2} is the network flow of the ring topology. There is a little difference for the ring topology comparing to the S1 case. The network flow of blockchain 1 is less than $1 \times 10^7$ bytes, which is similar to that in S1. However, those of blockchain 2 and 3 increase several times (maximum value approximate to $4 \times 10^7$ bytes), compared to the network flow in S1. The reason is that blockchain 3, which runs the PoS algorithm, has no need to use CPU to calculate a hash and has a higher capacity to process other tasks (including the transaction pre-processing task and the synchronization task). Its periodic tasks, which synchronize the blockchain data from its directly connected blockchain (blockchain 2), have more chances to be scheduled and to finish in time.

Another interesting point of the left part of Figure \ref{network_flow_s2} is that the sum of the network flow of blockchain 2 and 3 is more than the maximum network flow (of blockchain 1). It proves indirectly that no blockchain acts as the center to exchange network flow of all blockchains. This kind of topology has the potential to separate the network flow to each blockchain, which makes the scalability of this topology more flexible.

From the right part of Figure \ref{network_flow_s2}, we see the network flow of the router topology has almost the same tendency as in the S1 case. The network flow of blockchain 1 (the router blockchain) is the sum of another two blockchains. The sum and difference of blockchain 2 and blockchain 3 are shown in Figure \ref{network_flow_s2_router_sum_and_difference}.

\begin{figure*}
  \includegraphics[width=7in]{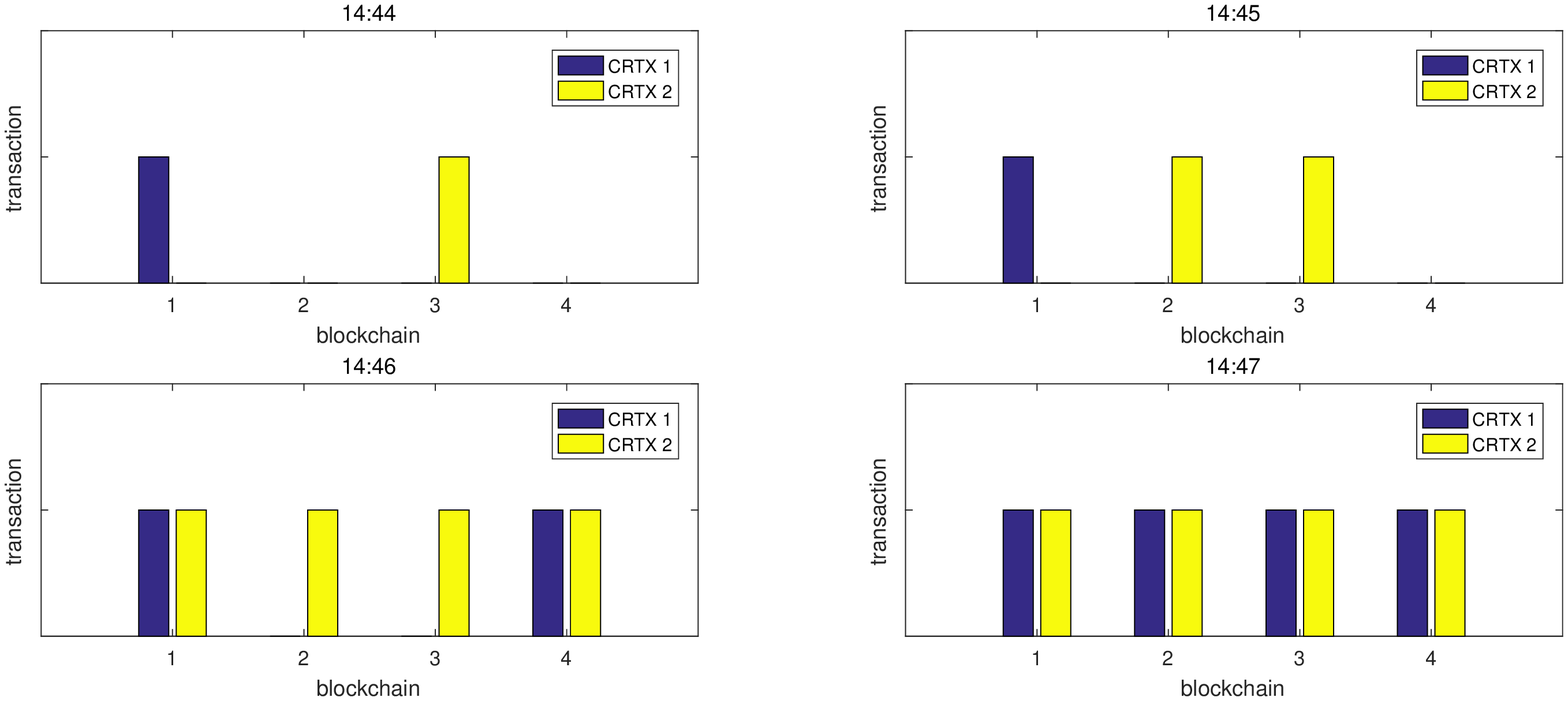}
  \caption{Ring topology with target transactions in indirectly connected blockchains. The time format is hh:mm, and this is the same in the following diagrams.}
  \label{transaction_propagation_1_3_indirect_connections}
\end{figure*}

\begin{figure*}
  \includegraphics[width=7in]{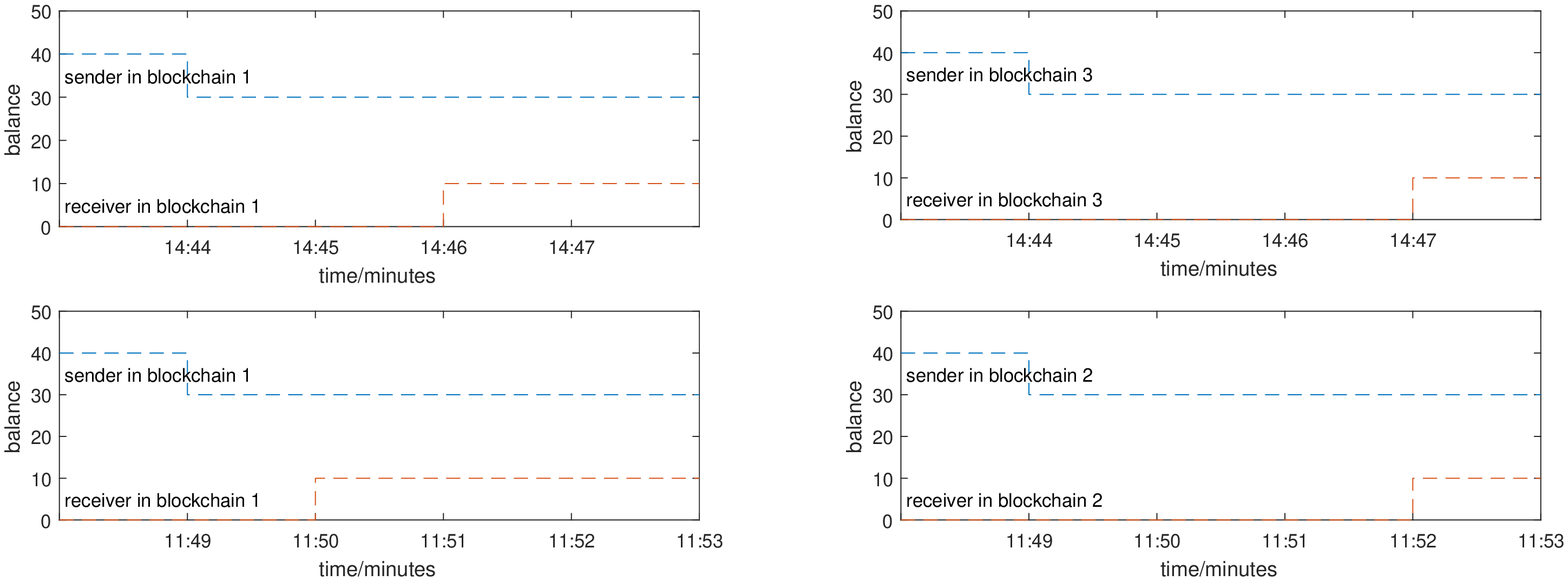}
  \caption{The balance of change process of the indirect case (diagrams at the upper part of this diagram) and the direct case (diagrams at the lower part of this diagram).}
  \label{transaction_propogation_balance_all}
\end{figure*}

Figure \ref{network_flow_s2_router_sum_and_difference} also shows the same tendencies as in the S1 case. The sum of the network flow of blockchain 1 and 2 is close or equal to that of router blockchain. The difference from S1 is the network flow difference among blockchain 1 and blockchain 2. The curve increases almost in a linearly way (in the result of a linear regression between the difference of network flow and the time, ‘multiple R’ is 0.9979, and ‘R Square’ is 0.9957).

In both S1 and S2 cases, we see different network flow distribution among the router topology and the router topology. In the router topology, the network flow is through the router blockchain. With the increase of connected blockchains, the burden of the router blockchain increases. However, the ring topology can dispatch the network flow to its directly connected blockchains, which indicates that it is a more flexible topology.

In the following sections, we try to verify the impact of different connection ways on the strongly connected topology and how the data validity is confirmed during the propagation process.

\subsection{Different Connection Way}
In this verification scenario, we describe the verification result with respect to how the cross-chain transactions are confirmed in the different connection ways, such as directly connected or indirectly connected. The consensus is PoW, as the result for PoS is similar, and we skip the result for PoW to save space.

\subsubsection{Cross-chain Interaction with Bridge Blockchains}
If the cross-chain interaction is among partial blockchains, the propagation among those blockchains needs the help of bridge blockchains. Bridge blockchains may occur in different positions, which causes other blockchains to form different connection ways, either directly connected or indirectly connected.

(1)Indirectly Connect

This verification is in a ring topology among blockchains from 1 to 4 (those blockchains are different from blockchains in the previous verification section as we reconstructed them), shown in Figure \ref{network1}. We want to verify the cross-chain data confirmation process among blockchain 1 and 3, which are indirectly connected through bridge blockchain 2 and 4. Two cross-chain transactions ($CRTX1$ and $CRTX2$) are used in the verification. $CRTX2$ is sent out in blockchain 3 and $CRTX1$ is sent out in blockchain 1. They are associated cross-chain transactions, which means the receiver of one transaction gets the asset when another transaction has been propagated into the former transaction’s blockchain.

Figure \ref{transaction_propagation_1_3_indirect_connections} shows the propagation result. Both transactions are carried out at almost the same time. As the mining periods are similar, both are sealed into its own blockchain at 11:44 (the time format is hh:mm, and the same is in the following). However, $CRTX2$ propagates more quickly as it is in a different propagation path. At 11:45, $CRTX2$ propagates to blockchain 2, and at 14:46 $CRTX2$ has been propagated to all blockchains. $CRTX1$ only propagates to blockchain 4 at 14:46. Finally, at time 14:47 $CRTX1$ propagates to all blockchains. Then, both transactions have been propagated to all blockchains no matter they are directly or indirectly connected.

We also show the balance change process during the propagation. The asset of a transaction transfers to its receiver when its required cross-chain transaction has been confirmed and propagated to the target blockchain. For example, if $CRTX2$ has been propagated to blockchain 1, the sender of $CRTX1$ can confirm that $CRTX2$ is in blockchain 3 and the asset is transferred to the receiver of $CRTX1$.

The balances of all accounts are dumped when any user’s balance has been changed. The balance change process is shown in the upper two parts of Figure \ref{transaction_propogation_balance_all}. Balances of senders in blockchain 1 and 3 are subtracted at time 14:44. Receivers get their assets at 11:46 and 11:47 separately when the associated transaction reaches. The time of receiver in blockchain 3 and 1 who receives the asset at a different time as the propagation is in different paths; one is $blockchain 1->blockchain 4->blockchain 3$, and another is $blockchain 3->blockchain 2->blockchain 1$.

\begin{figure}
    \includegraphics[width=2.8in]{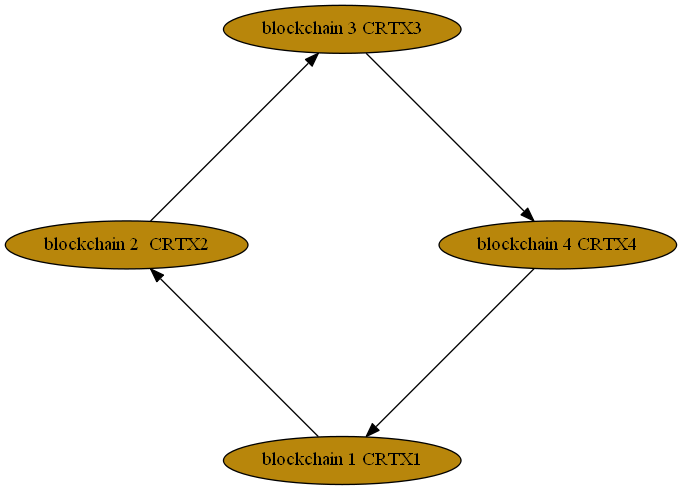}
    \caption{The topology of all directly connected blockchains.}
    \label{network_all_directly_neighbored}
  \end{figure}

\begin{figure*}
    \includegraphics[width=7.5in]{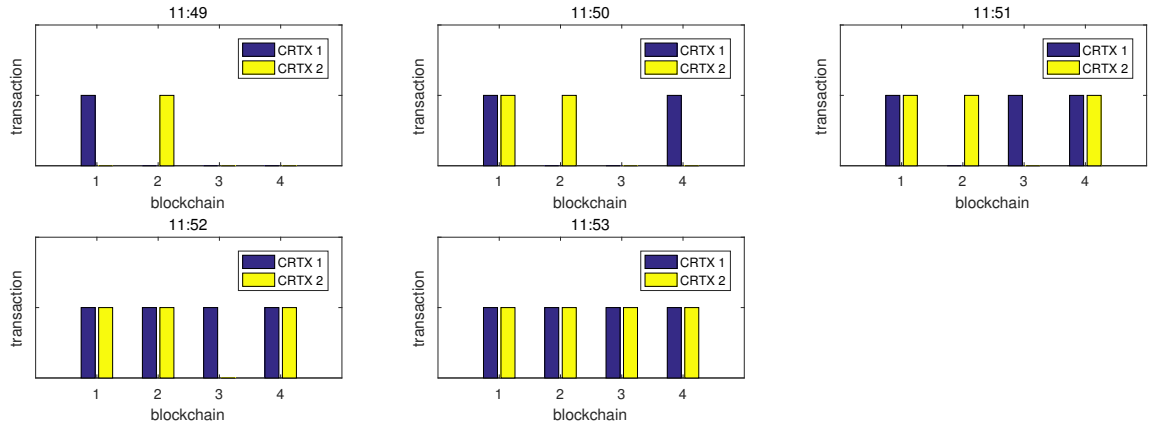}
    \caption{Ring topology with one target transaction in directly connected blockchains.}
    \label{transaction_propogation_neighbored_2}
  \end{figure*}

  \begin{figure}[htp]
    \includegraphics[width=3in]{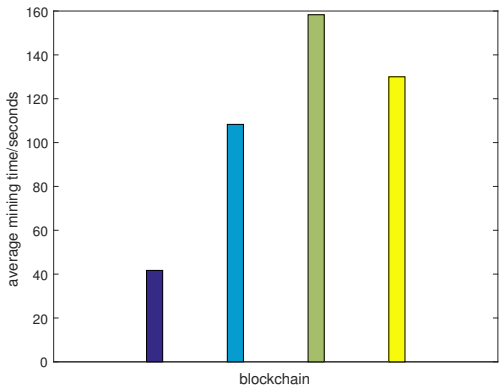}
    \caption{The average mining time of blockchain from 1 to 4}
    \label{mining_time_comparison}
  \end{figure}

(2) Cross-chain Interaction without Bridge Blockchains

In this verification, there are four cross-chain transactions $CRTX1$, $CRTX2$, $CRTX3$, and $CRTX4$. Each transaction is sent in one blockchain, and thus there are no bridge blockchains. Figure \ref{network_all_directly_neighbored} shows the topology of the blockchain connection which blockchain associated transactions.

 The propagation result is shown in Figure \ref{transaction_propogation_neighbored_2}. Two interesting points are here. The first one is that at 02:44 both $CRTx1$ and $CRTx2$ are sealed into blockchain 3. After we dump the blockchain data, both transactions are sealed into one block. This indicates that some transactions may be accumulated to one block due to different mining time and different mining overload.

The second issue is that blockchain 3 has the weakest calculation power. Its mining period is average longer than others, although the mining difficulties of all blockchains are the same (a newly mined block hash starts with 6 zeros). After checking the resources of its nodes, they are generally the weakest calculation resources (the fewest nodes and CPUs) for calculation. Figure \ref{mining_time_comparison} shows a comparison of those blockchain’s block mining time (in seconds).

\section{Related Work}
Cross-chain interactions aim to keep cross-chain exchanges atomic \cite{pre_16}\cite{pre_20}, which ensures that associated transactions are valid or invalid together. Current methods used to ensure atom swap include notary method, relay chain, or hash-locking \cite{pre_18}. Those methods check whether some target transactions are sealed in a specific blockchain or not. Cross-chain data is required to propagate in many cross-chain cases.

At the early stage, special topology is used in cross-chain interactions, such as in methods of relay-chain and pegged sidechain. The relay-chain method aims to solve the issue of extensibility and scalability of blockchains, and provide a scalable, heterogeneous multi-chain protocol\cite{pre_21}. This method is based on a star topology, in which the relay-chain connects all other parachains. Pegged sidechain originated from Bitcoin blockchain, and it enables the main chain (such as Bitcoin) and other blockchains (sidechain) to transfer assets among them \cite{pre_15}. The topology is among the main chain and its pegged chains. If there is one sidechain, it forms a mutually connected topology. If there are several sidechains, it is a star topology.

Later on,  some works \cite{pre_22}\cite{pre_23} propose the blockchain router, which aims to behave as the router of the Internet. It has a center ‘hub’ blockchain which connects all other blockchains. With the help of the center blockchain, other blockchains can exchange blockchain data. However, it also has disadvantaged as other blockchains are required to register and communicate with the center blockchain. It may burden the center blockchain with a heavy network flow, as all other blockchains communicate through it. Meanwhile, if we want to connect two centered blockchain networks to a new centered one, it requires all blockchains to connect to this new one.

For propagation methods, some methods rely on specific roles (such as validator, connector, surveillant, and nominator in \cite{pre_22}) or special accounts (such as escrow address in \cite{pre_23}) to keep the transactions atomic or data validity. In \cite{pre_23}, it adopts three-phase commit to synchronize the transaction state among blockchains. However, this method depends on a single node to communicate among blockchains, and it lacks a strong method to ensure data validation. Work \cite{pre_22} overcomes the disadvantages in which the validity of cross-chain data is not ensured. It adopts a special consensus algorithm (DS-PBFT) to confirm and keep the validity of cross-chain transactions.

The transaction format is also an interesting point. For heterogeneous blockchain, work\cite{pre_23} adopts a way to use the standard transaction data format \cite{pre_23}. Blockchains can interact with each other with this standard transaction format. The disadvantage is that all blockchains must adapt to this unified format, which is difficult when the account of associated blockchains are huge. For pegged blockchains \cite{pre_17}, the paired transactions appear in associated blockchains during the two-way peg process, in which a hidden transaction format transformation is done.

\section{Conclusions}
In this paper, we analyze the topology and the corresponding propagation method for the cross-chain transaction. It aims to provide a flexible blockchain topology with the data validity is ensured during the propagation. We point out the topology of blockchain is a strongly connected graph, and based on this principle we propose several topology instances for flexibility. Especially a ring topology model, which only requires a blockchain to connect to another one blockchain directly. It reduces the number of required neighbors. The propagation method is between directly connected neighbors, and it iterates to propagate the data among indirectly connected neighbors. With the proposed topology and propagation methods, the propagation of cross-chain transactions can be based on a flexible topology, and the data validity is kept. Future works include cross-chain interactions based on the propagated cross-chain transactions.

\appendix


%

\section*{Acknowledgment}

The authors thank the anonymous reviewers for their constructive comments, which help us to improve the quality of this paper. This work was supported in part by the National Natural Science Foundation of China under Grant No. 61772352; the Science and Technology Planning Project of Sichuan Province under Grant No. 2019YFG0400, 2018GZDZX0031, 2018GZDZX0004, 2017GZDZX0003, 2018JY0182, 19ZDYF1286.

\ifCLASSOPTIONcaptionsoff
  \newpage
\fi



%


\begin{IEEEbiographynophoto}
Su Hong received the BS and MS degrees, in 2003 and 2006, respectively, from Sichuan University, Chengdu, China.
He is currently a Ph.D. candidate in the School of Computer Science and software Sichuan University, Chengdu, China.
His research interests include blockchain and the Internet of Value.

\end{IEEEbiographynophoto}

\begin{IEEEbiographynophoto}
Guo Bing is currently a professor of Sichuan University, Ph.D. supervisor.
He received the Ph.D. from University of Electronic Science and Technology of China in 2002.
His current research interests include green computing, personal big data, and blockchain.    

\end{IEEEbiographynophoto}

\begin{IEEEbiographynophoto}
Shen Yan is currently a professor of Chengdu University of Information Technology. 
She received the Ph.D. degree from University of Electronic Science and Technology of China in 2004. 
Her research interests include smart terminal and instruments. 

\end{IEEEbiographynophoto}

\begin{IEEEbiographynophoto}
Tao Li is currently a professor of Sichuan University. 
His research interests include embedded computing and blockchain.

\end{IEEEbiographynophoto}

\end{document}